\documentclass[preprint,12pt]{elsarticle}

\usepackage{amsmath,amssymb}
\usepackage{graphicx}
\usepackage{wrapfig}

\usepackage{latexsym}

\usepackage[T2A]{fontenc}
\usepackage[cp1251]{inputenc}
\journal{Physica A: Statistical Mechanics and its Applications }

\begin{document}

\begin{frontmatter}

\address{1 Svientsitskii Street, 79011 Lviv, Ukraine}

\title{ Chain of kinetic equations for the distribution functions of particles in simple liquid taking into account nonlinear hydrodynamic fluctuations.}

\author{Petro~Hlushak and Mykhailo~Tokarchuk\\
        Institute for Condensed Matter Physics, Lviv, Ukraine\\
        }

\begin{abstract}
Chain of kinetic equations for non-equilibrium single, double and $s$-particle  distribution functions of particles is obtained taking into account nonlinear hydrodynamic fluctuations. Non-equilibrium distribution function of non-linear hydrodynamic fluctuations satisfies a generalized Fokker-Planck equation. The method of non-equilibrium statistical operator by Zubarev is applied.
A way of calculating of the structural distribution function of hydrodynamic collective variables and their hydrodynamic velocities (above Gaussian approximation) contained in the generalized Fokker-Planck equation for the non-equilibrium distribution function of hydrodynamic collective variables is proposed.
\end{abstract}


\end{frontmatter}

\section{Introduction   }
\label{}
The study of nonlinear kinetic and hydrodynamic fluctuations in dense gases,
liquids and plasma, in turbulence phenomena and dynamics of phase transitions,
in chemical reactions and self-organizing processes are relevant both on kinetic
and hydrodynamic levels of description in statistical theory of non-equilibrium processes
\cite{Tokuyama2014,RanJuil2014,ValenVass2013,Klimontov2002,MendLopVizc2014,Juilin2012,LangShillKrFr2012,
YoshArim2013,BoonLutsLuts2012,Mazenko2010,Mazenko2011,Mazenko2013,KostTokTokMark2014,HlushTok2014,
SilVasRamLuz2011,SilVasRamLuzPr2012,TsytAnd2004,Olemskoi1999,Frank2004,MarkTokKostTok2011}.
The non-equilibrium states of such systems are far from equilibrium.
Therefore the study of both the processes establishing the stationary states
with characteristic times of life and the relaxation processes to the known equilibrium states,
that are described by means of molecular hydrodynamics \cite{BoonYip1980,MrygTok1992,MrygOmelTok1995,MarkOmelTok2010,ZubMorOmTok1993},
is of great importance.
An important feature of theoretical modeling of
non-equilibrium phenomena in dense gases, liquids, dense plasmas (dusty plasmas) is a consistent description
of kinetic and hydrodynamic processes \cite{ZubMorOmTok1993,TokOmelKobr1998,KobrOmelTok1998,MarkOmelTok2014,MarkTok2014}
and taking into account the characteristic short and long-range interactions
between the particles of the systems.
In particular, the non-equilibrium gas-liquid phase transition
is characterized by nonlinear hydrodynamic fluctuations of mass,
momentum and  particle energy, which describe a collective nature
of the process and define the spatial and temporal
behavior of the transport coefficients (viscosity, thermal
conductivity), time correlation functions and dynamic structure factor.
At the same time, due to heterogeneity in collective dynamics
of these fluctuations, liquid drops emerge in the gas phase
(in case of transition from the gas phase to the liquid phase),
or the gas bubbles emerge in the liquid phase
(in case of transition from the liquid phase to the gas phase),
formation of which has a kinetic nature described by a
redistribution of momentum and energy, i.e.
when a certain group of particles in the system receives a significant
decrease (in the case of drops), or increase (in the case of bubbles)
of kinetic energy.
The particles, that form bubbles or droplets, diffuse out of their phases
in the liquid or the gas and vice versa. They have different values
of momentum, energy and pressure in different phases. All these
features are related to the non-equilibrium unary, binary and $s$-particle
distribution functions (which depend on coordinate, momentum and time)
that satisfy the BBGKY chain of equations.
These problems concern the consistent description of kinetic and hydrodynamic processes
in heterophase systems \cite{Yukalov1991,OlemKopl1995,Onuki2004,Das2011}.

Therefore, the construction of kinetic equations that take into account
nonlinear hydrodynamic fluctuations \cite{Bogolub1978,Dorfman1981,Klimontov1992:1,Klimontov1992:2,Cogen1993}
is an important problem in the theory of transport processes in dense
gases and liquids. In particular, this problem arises in the description
of low-frequency anomalies in the kinetic equations and related "long
tail" correlation functions \cite{ResibLeen1977,SchnHoflHoflVoig2011,Franos2014}.

The main difficulty of the problem is that the kinetics and hydrodynamics
of these processes are strongly related and should be considered simultaneously.
Zubarev, Morozov, Omelyan and Tokarchuk \cite{ZubMor1984,ZubMorOmTok1991,ZubMorOmTok1993}
proposed the consistent
description of kinetic and hydrodynamic processes in dense gases and liquids
on the basis of Zubarev non-equilibrium statistical operator
\cite{ZubMorRop1996:1,ZubMorRop1996:2}.
In particular, this approach was used to obtain
the kinetic equation of the revised Enskog theory \cite{ZubMorOmTok1991,KobMorOmTok1996}
for a system of hard spheres
and kinetic Enskog-Landau equations for one-component system of
charged hard spheres from the BBGKY chain of equations.

Zubarev  {\it et al} \cite{ZubMorOmTok1993} obtained
the generalized hydrodynamic equations for
the hydrodynamic variables (densities of the particle,
momentum and the total energy) connected with the kinetic
equation for the non-equilibrium one-particle distribution
function. Later \cite{TokOmelKobr1998,MarkOmelTok2014},
these equations were used to study time correlation functions
and the collective excitation spectrum of the weakly
non-equilibrium processes in liquids.

Obviously, the approach proposed by Zubarev  {\it et al} \cite{ZubMor1984,ZubMorOmTok1991,ZubMorOmTok1993}
and Tokarchuk {\it et al} \cite{TokOmelKobr1998,MarkOmelTok2014} can be used to describe
both weakly and strongly non-equilibrium systems. At the same time,
in order to consistently describe kinetic processes and nonlinear
hydrodynamic fluctuations it is convenient to reformulate this theory so that
a set of equations for non-equilibrium one-particle distribution function
and the distribution functional of hydrodynamic variables,
particle number densities as well as momentum and energy densities could be obtained.

In this contribution we will develop an approach for consistent description
of kinetic and hydrodynamic processes that are characterized by non-linear
fluctuations and are especially important for the description of
non-equilibrium gas-liquid phase transition.
In the second section we will obtain the non-equilibrium statistical operator
for non-equilibrium state of the system when the parameters of the reduced
description are a non-equilib\-rium one-particle distribution function and
the distribution function of non-equilibrium nonlinear hydrodynamic variables.
Using this operator we construct the kinetic equations for the non-equilibrium
single, double, $s$-particle distribution functions which take into account
nonlinear hydrodynamic fluctuations, for which the non-equilibrium distribution
function satisfies a generalized Fokker-Planck equation. In the third section,
we will consider one of the ways to calculate the structural distribution function
of hydrodynamic collective variables and their hydrodynamic velocities
(in higher than Gaussian approximation), which enter the generalized Fokker-Planck
equation for the non-equilibrium distribution function of hydrodynamic
collective variables.

\section{Non-equilibrium distribution function}

For a consistent description of kinetic and hydrodynamic fluctuations in a classical one-component
fluid it is necessary to select the description parameters for one-particle and collective
processes. As these parameters we choose the non-equilibrium one-particle distribution function
$f_{1}(x;t)=\langle\hat{n}_{1}(x)\rangle^{t}$ and distribution function of hydrodynamic variables
$f(a;t)=\langle\delta(\hat{a}-a)\rangle^{t}$. Here the phase function

\begin{equation}    \label{Eq.1}
\hat{n}_{1}(x)=\sum_{j=1}^{N}\delta(x-x_{j})=
\sum_{j=1}^{N}\delta({\bf r}-{\bf r}_{j})\delta({\bf p}-{\bf p}_{j})
\end{equation}
is the microscopic particle number density. $x_{j}=({\bf r}_{j},{\bf p}_{j})$ is the set of
phase variables (coordinates and momentums), $N$ is the total number of particles in a volume $V$.
A microscopic phase distribution of hydrodynamic variables is given by
\begin{equation}    \label{Eq.2}
\hat{f}(a)=\delta (\hat{a}-a)=
\prod_{m=1}^{3}\prod_{\bf {k}}\delta (\hat{a}_{m \bf {k}}-a_{m \bf {k}}),
\end{equation}
where $\hat{a}_{1 \bf {k}}=\hat{n}_{\bf {k}}$, $\hat{a}_{2\bf{k}}=\hat{\bf{J}}_{\bf{k}}$,
$\hat{a}_{3 \bf{k}}=\hat{\varepsilon}_{\bf {k}}$ are the Fourier components of the densities of
particle number, momentum and energy:
\begin{eqnarray}    \label{Eq.3}
\hat{n}_{\bf k}=\sum_{j=1}^{N}\mbox{e}^{-i{\bf k} {\bf r}_{j}},  \qquad
\hat{\bf J}_{\bf k}=\sum_{j=1}^{N}{\bf p}_{j}\mbox{e}^{-i{\bf k} {\bf r}_{j}},  \\
\nonumber
\hat{\varepsilon}_{\bf k}= \sum_{j=1}^{N}\Big[ \frac{p^{2}_{j}}{2m}+\frac{1}{2}\sum_{l\neq
j=1}^{N}\Phi(|{\bf r}_{l j}|)\Big]
 \mbox{e}^{-i{\bf k} {\bf r}_{j}},
\end{eqnarray}
and $a_{m{\bf k}}=(n_{{\bf k}},{\bf J}_{\bf{k}},\varepsilon_{\bf k})$ are the corresponding
collective variables. $\Phi(|{\bf r}_{lj}|)=\Phi(|{\bf r}_{l}-{\bf r}_{j}|)$ is the pair
interaction potential between particles. The average values $\langle\hat{n}_{1}(x)\rangle^{t}$ and
$\langle\delta(\hat{a}-a)\rangle^{t}$ are calculated by means of the non-equilibrium $N$-particle
distribution function $\varrho(x^{N};t)$, that satisfies the Liouville equation. In line with
the idea of reduced description of non-equilibrium states this function is the functional
\begin{equation}    \label{Eq.4}
\varrho (x^{N};t)=\varrho (\ldots,f_{1}(x;t),f(a;t),\ldots).
\end{equation}

In order to find a non-equilibrium distribution function $\varrho(x^{N};t)$ we use
Zubarev's method \cite{ZubMorRop1996:1,ZubMorRop1996:2,ZubMor1989}, in which
a general solution of Liouville equation taking into account a projection procedure
can be presented in the form:
\begin{equation}    \label{Eq.5}
\varrho(x^{N};t)=\varrho_{rel}(x^{N};t)-\int_{-\infty}^{t}dt'
  \mbox{e}^{\epsilon(t'-t)}T(t,t')(1-P_{rel}(t'))iL_{N}\varrho_{rel}(x^{N};t'),
\end{equation}
where $\epsilon\rightarrow +0$ after thermodynamic limiting transition. The source selects the
retarded solutions of Liouville equation with operator $iL_{N}$.
$T(t,t')=\exp_{+}(-\int_{t'}^{t}dt'(1-P_{rel}(t'))iL_{N})$ is the generalized time evolution
operator taking into account Kawasaki-Gunton projection $P_{rel}(t')$. The structure of $P_{rel}(t')$
depends on the relevant distribution function $\varrho_{rel}(x^{N};t)$, which in method
by Zubarev is determined from extremum of the information entropy at simultaneous conservation of
normalization condition
\begin{equation}    \label{Eq.6}
 \int d\Gamma_{N}\varrho_{rel} (x^{N};t)=1,\,\,\, d\Gamma_{N}=\frac{(dx)^{N}}{N!}=\frac{(dx_{1},\,\,\, \ldots,dx_{N})}{N!},\,\,\,
  dx=d{\bf r} d{\bf p},
\end{equation}
and the fact that the parameters of the reduced description,
$f_{1}(x;t)$ and $f(a;t)$ are fixed.
Then relevant distribution function can be written as
\begin{equation}    \label{Eq.7}
 \varrho_{rel} (x^{N};t)=\exp \Big\{-\Phi(t)-\int dx \gamma(x;t)\hat{n}_{1}(x)-\int da F(a;t)\hat{f}(a)\Big\},
\end{equation}
where $da$ is the integration over collective variables:
\begin{eqnarray}     \nonumber
da = \prod\limits_{\bf k}dn_{\bf k}d{\bf j}_{\bf k}d\varepsilon_{\bf k}, \,\,
  dn_{\bf k}=d\,\mbox{Re} n_{\bf k}\,d\,\mbox{Im}n_{\bf k},       \,\,
\varepsilon_{\bf k} = d\,\mbox{Re} \varepsilon_{\bf k}\,d\,
                            \mbox{Im} \varepsilon_{\bf k},       \\  \nonumber
d{\bf j}_{\bf k} = d j_{x,{\bf k}}\,d j_{y,{\bf k}}d j_{z,{\bf k}},
 \quad
d j_{x,{\bf k}}=d\,\mbox{Re}  j_{x,{\bf k}}\,d\,\mbox{Im} j_{x,{\bf k}},\ldots .
\end{eqnarray}

The Massieu-Planck functional $\Phi(t)$ is determined from the normalization condition
for the relevant distribution function
$$
 \Phi (t)=\ln \int d\Gamma_{N}\exp \Big\{-\int dx \gamma (x;t)\hat{n}_{1}(x)-\int da F(a;t)\hat{f}(a)\Big\}.
$$
The functions $\gamma(x;t)$ and $F(a;t)$ are the Lagrange multipliers and are determined
from the self-consistent conditions:
\begin{equation}    \label{Eq.8}
 f_{1}(x;t)=\langle \hat{n}_{1}(x) \rangle^{t}=\langle \hat{n}_{1}(x) \rangle^{t}_{rel}, \quad
  f(a;t)=\langle \delta (\hat{a}-a) \rangle^{t}=\langle \delta (\hat{a}-a) \rangle^{t}_{rel},
\end{equation}
where $\langle \ldots \rangle^{t}=\int d\Gamma_{N} \ldots \varrho(x^{N};t)$ and
$\langle \ldots \rangle^{t}_{rel}=\int d\Gamma_{N} \ldots \varrho_{rel}(x^{N};t)$.
To find the explicit form of non-equilibrium distribution function $\varrho(x^{N};t)$
we exclude the factor $F(a;t)$ in relevant distribution function and thereafter,
by means of self-consistent conditions (\ref{Eq.8}), we have
\begin{equation}    \label{Eq.9}
 \varrho_{rel} (x^{N};t)=\varrho_{rel}^{kin} (x^{N};t)\frac{f(a;t)}{W(a;t)}\Big{|}_{a=\hat{a}}.
\end{equation}
Here
\begin{eqnarray}    \label{Eq.10}
 W(a;t)=\int d\Gamma_{N}\mbox{e}^{-\Phi^{kin}(t)- \int dx \gamma(x;t)\hat{n}_{1}(x)}\hat{f}(a)=
   \int d\Gamma_{N}\varrho_{rel}^{kin}(x^{N};t)\hat{f}(a)
\end{eqnarray}
is the structure distribution function of hydrodynamic variables, which
could be also considered as
a Jacobian for transition from $\hat{f}(a)$ into space of collective variables
$n_{\bf{k}}$, $\bf{J}_{\bf{k}}$, $\varepsilon_{\bf{k}}$ averaged with the "kinetic"
relevant distribution function
\begin{eqnarray}    \label{Eq.11}
 \varrho_{rel}^{kin} (x^{N};t)=\exp\Big\{-\Phi^{kin}(t)-\int dx\gamma(x;t)\hat{n}_{1}(x)\Big\}, \\
\nonumber
 \Phi^{kin}(t)=\ln \int d\Gamma_{N}\exp \Big\{-\int dx \gamma (x;t)\hat{n}_{1}(x)\Big\}.
\end{eqnarray}
Here the entropy
\begin{eqnarray}    \label{Eq.12}
 S(t)=-\langle \ln \varrho_{rel} (x^{N};t)  \rangle^{t}_{rel}   \qquad\qquad        \\
\nonumber
  =\Phi (t)+\int dx \gamma (x;t)\langle \hat{n}_{1}(x)\rangle^{t}+\int da f(a;t)\ln \frac{f(a;t)}{W(a;t)}.
\end{eqnarray}
corresponds to the relevant distribution (\ref{Eq.9}).
In combination with the self-consistent conditions (\ref{Eq.8}), it can be considered as
entropy of non-equilibrium state. In accordance with (\ref{Eq.5}), in order to obtain
the explicit form of non-equilibrium distribution function, it is necessary to disclose the action
of Liouville operator on $\varrho_{rel}(x^{N};t)$ and action of the Kawasaki-Gunton
projection operator, which in our case has the following structure according to (\ref{Eq.9}):
\begin{eqnarray}    \label{Eq.13}
 P_{rel}(t)\varrho'=\varrho_{rel} (x^{N};t)\int d\Gamma_{N}\varrho'+
\int dx \frac{\partial\varrho_{rel}(x^{N};t)}{\partial\langle\hat{n}_{1}(x)\rangle
  \mbox{e}^{t}}         \qquad\qquad                                      \\
\nonumber
 \times\Big(\int d\Gamma_{N}\hat{n}_{1}(x)\varrho'-
  \langle \hat{n}_{1}(x) \rangle^{t}\int d\Gamma_{N}\varrho'\Big)    \qquad  \\
\nonumber
+\int da \frac{\partial \varrho_{rel} (x^{N};t)}{\partial (\frac{f(a;t)}{W(a;t)})}\frac{1}{W(a;t)}\Big(\int d\Gamma_{N}\hat{f}(a)\varrho'-
f(a;t)\int d\Gamma_{N}\varrho'\Big)                           \\
\nonumber
+\int dx \int da \frac{\partial \varrho_{rel} (x^{N};t)}{\partial (\frac{f(a;t)}{W(a;t)})}\frac{f(a;t)}{W(a;t)}\frac{\partial \ln W(a;t)}{\partial \langle \hat{n}_{1}(x) \rangle^{t}}           \qquad\qquad                   \\
\nonumber
\times\Big(\int d\Gamma_{N}\hat{n}_{1}(x)\varrho'-
\langle \hat{n}_{1}(x) \rangle^{t}\int d\Gamma_{N}\varrho'\Big).   \qquad\qquad
\end{eqnarray}
Next, we consider the action of Liouville operator on relevant distribution
function (\ref{Eq.9}):
\begin{eqnarray}    \label{Eq.14}
iL_{N}\varrho_{rel} (x^{N};t)=-\int dx \gamma (x;t)\dot{\hat{n}}_{1}(x)\varrho_{rel} (x^{N};t)  \\
\nonumber
+\left[iL_{N}\frac{f(a;t)}{W(a;t)}\Big{|}_{a=\hat{a}} \right] \varrho_{rel}^{kin} (x^{N};t),
\end{eqnarray}
where $\dot{\hat{n}}_{1}(x)=iL_{N}\hat{n}_{1}(x)$. Having used thereafter the relation
\begin{eqnarray}
\nonumber
iL_{N}\hat{f}(a)=iL_{N}\hat{f}(n_{\bf k},{\bf J}_{\bf k},\varepsilon_{\bf {k}})  \qquad\qquad  \\
\nonumber
=\sum_{\bf k} \Big[\frac {\partial }{\partial n_{\bf {k}}}\hat{f}(a) \dot{\hat{n}}_{\bf {k}}+
\frac {\partial }{\partial {\bf J}_{\bf k}}\hat{f}(a) \dot{\hat{ {\bf J}}}_{\bf k}+
\frac {\partial }{\partial \varepsilon_{\bf  k}} \hat {f}(a) \dot{\hat{\varepsilon}}_{\bf {k}}
\Big],
\end{eqnarray}
where $\dot{\hat{n}}_{\bf k}=iL_{N}\hat{n}_{\bf k},\quad
\dot{\hat{{\bf J}}}_{\bf  k}=iL_{N}\hat{{\bf J}}_{\bf k}, \quad
\dot{\hat{\varepsilon}}_{\bf k}=iL_{N}\hat{\varepsilon}_{\bf k}$,
the last expression in (\ref{Eq.14}) can be rewritten in following form:
\begin{eqnarray}    \label{Eq.15}
\left[iL_{N}\frac{f(a;t)}{W(a;t)}\Big{|}_{a=\hat{a}}\right] \varrho_{rel}^{kin} (x^{N};t)=
\int da \sum_{\bf k} W(a;t) \Big[\dot{\hat{n}}_{\bf k}
\frac {\partial }{\partial n_{\bf k}} \frac {f(a;t)}{W(a;t)}      \\
\nonumber
+\,\dot{\hat{{\bf J}}}_{\bf k}\frac{\partial}{\partial{\bf J}_{\bf k}}\frac{f(a;t)}{W(a;t)}+
\dot{\hat{\varepsilon}}_{\bf k} \frac{\partial}{\partial\varepsilon_{\bf k}}\frac{f(a;t)}
{W(a;t)}\Big]\varrho_{L}(x^{N};t).          \qquad\qquad
\end{eqnarray}
Here we introduced new relevant distribution function $\varrho_{L} (x^{N},a;t)$ with the
microscopic distribution of large-scale collective variables
\begin{equation}    \label{Eq.16}
\varrho_{L} (x^{N},a;t)=\varrho_{rel}^{kin} (x^{N};t)  \frac{\hat{f}(a)}{ W(a;t)}.
\end{equation}
This relevant distribution function is connected with $\varrho_{rel}(x^{N};t)$
by the relation
\begin{equation}    \label{Eq.17}
\varrho_{rel}(x^{N};t)=\int daf(a;t)\varrho_{L}(x^{N},a;t)
\end{equation}
and is obviously normalized to unity
\begin{equation}    \label{Eq.18}
\int d\Gamma_{N}\varrho_{L}(x^{N},a;t)=1.
\end{equation}
Using then the relation (\ref{Eq.16}), the average values with relevant distribution
is convenient to represent in following form:
\begin{equation}   \label{Eq.19}
 \langle \ldots \rangle^t_q = \int da \langle \ldots \rangle^t_L f(a;t),\quad
 \langle \ldots \rangle^t_L = \int d\Gamma_N \ldots\varrho_L(x^N,a;t).
\end{equation}
Now in accordance with (\ref{Eq.15}) and (\ref{Eq.16}) we can rewrite the action of the Liouville
operator on $\varrho_{rel}(x^N;t)$ as follows
\begin{eqnarray}    \label{Eq.20}
 iL_N \varrho_{rel}(x^N;t)=-\int da \int dx \gamma(x;t)\dot{\hat{n}}_1(x)\varrho_{L}(x^N,a;t)f(a;t) \\
\nonumber
+\int da \sum\limits_{\bf{k}}  W(a;t)
  \Big[\dot{\hat{n}}_{\bf k}\frac{\partial}{\partial n_{\bf k}}\frac{f(a;t)}{W(a;t)}+
  \dot{\hat{{\bf J}}}_{\bf k}\cdot\frac{\partial}{\partial{\bf J}_{\bf k}}
  \frac{f(a;t)}{W(a;t)}                                             \\
\nonumber
  +\dot{\hat{\varepsilon}}_{\bf k}\frac{\partial}{\partial\varepsilon_{\bf k}}
  \frac{f(a;t)}{W(a;t)}\Big]\varrho_{L}(x^{N},a;t).   \qquad  \qquad
\end{eqnarray}
Substituting this expression into (\ref{Eq.5}), one obtains for non-equilibrium
distribution function the following result:
\begin{eqnarray}    \label{Eq.21}
\varrho (x^{N};t)=\int da f(a;t)\varrho_{L}(x^{N},a;t)    \qquad\qquad \qquad\qquad   \\
\nonumber
+\int da \int dx  \int_{-\infty}^{t}dt'e^{\epsilon (t'-t)}T(t,t')\Big(1-P_{rel}(t')\Big)
  \qquad\qquad \qquad \\
\nonumber
\times \dot{\hat{n}}_{1}(x)\varrho_{L} (x^{N},a;t')f(a;t')\gamma (x;t')   \qquad\qquad \qquad\qquad\\
\nonumber
-\int da \sum_{\bf k}  \int_{-\infty}^{t}dt'e^{\epsilon (t'-t)}T(t,t')\Big(1-P_{rel}(t')\Big)
W(a;t')\Big[\dot{\hat{n}}_{\bf k}\frac{\partial}{\partial n_{\bf k}}\frac{f(a;t')}{W(a;t')}  \\
\nonumber
+\dot{\hat{{\bf J}}}_{\bf k}\cdot\frac{\partial}{\partial{\bf J}_{\bf k}}
\frac{f(a;t')}{W(a;t')}+ \dot{\hat{\varepsilon}}_{\bf k}
\frac{\partial}{\partial\varepsilon_{\bf k}}
\frac{f(a;t')}{W(a;t')}\Big]\varrho_{L}(x^{N},a;t').  \qquad
\end{eqnarray}
and the corresponding generalized transport equations:
\begin{eqnarray}    \label{Eq.22}
\nonumber
\Big[\frac{\partial}{\partial t}+\frac{\bf p}{m}\cdot\frac{\partial}{\partial \bf r}\Big]
f_{1}(x;t)-\int dx'\frac{\partial }{\partial \bf r}\Phi(|{\bf r} -{\bf r}'|)
\cdot\Big[\frac{\partial }{\partial \bf p}-
\frac{\partial }{\partial {\bf p}'}\Big]g_{2}(x,x';t) \\
=\int dx'\int da\int_{-\infty}^{t}dt'e^{\epsilon(t'-t)}\phi_{nn}(x,x',a;t,t')f(a;t')\gamma (x';t')\\
\nonumber
-\sum_{\bf k}\int da\int_{-\infty}^{t}dt'e^{\epsilon (t'-t)}
\Big\{\phi_{nj}(x,{\bf k},a;t,t')\cdot \frac {\partial  }{\partial {\bf J}_{\bf k}}   \qquad\qquad\\
\nonumber
+\phi_{n\varepsilon}(x,{\bf k},a;t,t')\frac{\partial}{\partial\varepsilon_{\bf k}}\Big\}
\frac{f(a;t')}{W(a;t')},           \qquad\qquad \qquad\qquad
\end{eqnarray}
\begin{eqnarray}    \label{Eq.23}
\nonumber
\frac{\partial }{\partial t}f(a;t)=\sum_{\bf k}
\Big\{\frac {\partial }{\partial n_{{\bf k}}}v_{n,{\bf k}}(a;t)+
\frac {\partial  }{\partial {\bf J}_{\bf k}}\cdot {\bf v}_{j,{\bf k}}(a;t)+
\frac {\partial  }{\partial \varepsilon_{\bf k}}v_{\varepsilon,{\bf k}}(a;t)\Big\}f(a;t) \qquad\\
\nonumber
=\sum_{\bf k}\frac {\partial  }{\partial {\bf J}_{\bf k}}\cdot\int dx'\int da'\int_{-\infty}^{t}dt'
e^{\epsilon (t'-t)}\phi_{jn}(x',{\bf k},a,a';t,t')f(a;t')\gamma (x';t')  \\
-\sum_{\bf k}\frac {\partial  }{\partial \varepsilon_{\bf k}}\int dx'\int da'\int_{-\infty}^{t}dt'
e^{\epsilon(t'-t)}\phi_{\varepsilon n}(x',{\bf k},a,a';t,t')f(a;t')\gamma (x';t') \quad\\
\nonumber
+\sum_{\bf k,\bf q}\int da'\int_{-\infty}^{t}dt' e^{\epsilon(t'-t)}\frac{\partial}{\partial
{\bf J}_{\bf k}}\cdot \phi_{jj}({\bf k},{\bf q},a,a';t,t')
\cdot\frac {\partial  }{\partial {\bf J}_{\bf q}}\frac {f(a;t')}{W(a;t')}  \qquad \\
\nonumber
+\sum_{\bf k,\bf q}\int da'\int_{-\infty}^{t}dt' e^{\epsilon (t'-t)}
\frac {\partial}{\partial {\varepsilon}_{\bf k}}
\phi_{\varepsilon \varepsilon}({\bf k},{\bf q},a,a';t,t')
\frac {\partial}{\partial {\varepsilon}_{\bf q}}\frac {f(a;t')}{W(a;t')}  \qquad\qquad\\
\nonumber
+\sum_{\bf k,\bf q}\int da'\int_{-\infty}^{t}dt' e^{\epsilon (t'-t)}
\Big\{\frac{\partial}{\partial{\bf J}_{\bf k}}\cdot\phi_{j\varepsilon}({\bf k},{\bf q},a,a';t,t')
\frac {\partial}{\partial{\varepsilon}_{\bf q}}   \qquad\qquad \\
\nonumber
+\frac{\partial}{\partial{\varepsilon}_{\bf k}}
\phi_{\varepsilon j}({\bf k},{\bf q},a,a';t,t')\cdot\frac{\partial}{\partial{\bf J}_{\bf q}}\Big\}
\frac {f(a;t')}{W(a;t')}. \qquad\qquad \qquad\qquad
\end{eqnarray}
The generalized transport equations (\ref{Eq.22}), (\ref{Eq.23}) include
the relevant binary distribution function of particles $g_{2}(x,x';t)$ :
\begin{eqnarray}    \label{Eq.24}
g_{2}(x,x';t)= \langle G_2(x,x')\rangle_q^{t}
=\langle\hat{n}_{1}(x)\hat{n}_{1}(x')\rangle_q^{t}            \\
=\int d\Gamma_{N-2}\varrho_q(x^{N};t)=
\int da g_{2}^{L}(x,x';a;t)f(a;t),         \nonumber
\end{eqnarray}
where
$$
g_{2}^{L}(x,x';a;t)=\int d\Gamma_{N-2}\varrho_{L} (x^{N};a;t)
$$
is the binary relevant distribution function of large-scale collective variables.
The generalized transport kernels $\phi_{\alpha \beta}$ ($\alpha, \beta =\{n,{\bf{j}},\varepsilon\}$),
that describe non-Markovian kinetic and hydrodynamic processes, are non-equilibrium correlation
functions of generalized fluxes $I_{\alpha}, I_{\beta}$:
\begin{equation}    \label{Eq.25}
\phi_{\alpha \beta}(t,t')=\langle I_{\alpha}(t)T(t,t')I_{\beta}(t') \rangle_{L}^{t'},
\end{equation}
\begin{equation}    \label{Eq.26}
\hat{I}_{n}(x;t)=\big(1-P(t)\big)\dot{\hat{n}}_{1}(x),
\end{equation}
\begin{equation}    \label{Eq.27}
\hat{I}_{\bf j}({\bf k};t)=\big(1-P(t)\big)\dot{\hat{\bf J}}_{\bf k},
\end{equation}
\begin{equation}    \label{Eq.28}
\hat{I}_{\varepsilon}({\bf k};t)=\big(1-P(t)\big)\dot{\hat{\varepsilon}}_{\bf k}.
\end{equation}
Here $P(t)$ is the generalized Mori operator related to Kawasaki-Gunton projection operator
$P_{rel}(t)$ by following relation
$$
P_{rel}(t)a(x)\varrho_{rel} (x^{N};t)=\varrho_{rel} (x^{N};t)P(t)a(x).
$$
It should be emphasized that in (\ref{Eq.25}) the averages are calculated with
distribution function $\varrho_{L}(x^{N},a;t)$ (\ref{Eq.19}), so that the transport kernels
are some functions of collective variables $a_{\bf k}$.
In equation (\ref{Eq.23}) the functions (called hydrodynamic velocities) $v_{n,{\bf k}}(a;t)$, ${\bf v}_{j,{\bf k}}(a;t)$, $v_{\varepsilon,{\bf k}}(a;t)$ represent the fluxes
in the space of collective variables and are defined as:
\begin{eqnarray}        \label{Eq.29}
\nonumber
 v_{n,{\bf k}}(a;t)=\int d\Gamma_{N}\dot{\hat{n}}_{\bf k}\varrho_{L} (x^{N},a;t)= \langle
\dot{\hat{n}}_{\bf k} \rangle_{L}^{t},                \\
 {\bf v}_{j,{\bf k}}(a;t)=\int d\Gamma_{N}\dot{\hat{{\bf J}}}_{\bf k}\varrho_{L} (x^{N},a;t)= \langle
 \dot{\hat{{\bf J}}}_{\bf k} \rangle_{L}^{t},         \\
\nonumber
 v_{\varepsilon,{\bf k}}(a;t)=\int d\Gamma_{N}\dot{\hat{\varepsilon}}_{\bf k}\varrho_{L} (x^{N},a;t)= \langle \dot{\hat{\varepsilon}}_{\bf k} \rangle_{L}^{t}.
\end{eqnarray}

The presented system of transport equations gives consistent description of kinetic and
hydrodynamic processes of classical fluids which take into account long-living fluctuations.

The system of transport equations (\ref{Eq.22}), (\ref{Eq.23}) is a not closed
due to Lagrange parameter $\gamma(x;t)$,
which is determined from the corresponding self-consistent conditions.
From the kinetic processes standpoint,  we must supplement this system of transport equations
with the kinetic equation $f_{2}(x,x';t)$, and hence for $f_{s}(x_1 \dots x_s;t)$,  $s<N$:
\begin{eqnarray}    \label{Eq.30}
\nonumber
\frac{\partial}{\partial t}
f_{2}(x,x';t)+iL_{2}f_{2}(x,x';t)-\int dx''\{i L(x,x'')+i L(x',x'')\}f_{3}(x,x',x'';t) \\
\nonumber=iL_{2}\Delta f_{2}(x,x';t)-\int dx''\{i L(x,x'')+i L(x',x'')\}\Delta f_{3}(x,x',x'';t) \qquad\\
+\int dx''\int da\int_{-\infty}^{t}dt'e^{\epsilon(t'-t)}\phi_{Gn}(x,x',x'',a;t,t')f(a;t')\gamma (x'';t')  \qquad\\
\nonumber
-\sum_{\bf k}\int da\int_{-\infty}^{t}dt'e^{\epsilon (t'-t)}
\Big\{\phi_{Gj}(x,x',{\bf k},a;t,t')\cdot \frac {\partial  }{\partial {\bf J}_{\bf k}} \qquad\qquad  \\
+ \,\phi_{G\varepsilon}(x,x',{\bf k},a;t,t')\frac{\partial}{\partial\varepsilon_{\bf k}}\Big\}
\frac{f(a;t')}{W(a;t')},\nonumber \qquad\qquad \qquad\qquad
\end{eqnarray}

\begin{eqnarray}    \label{Eq.31}
\frac{\partial}{\partial t}f_{s}(x_1\dots x_s;t)+iL_{s}f_{s}(x_1\dots x_s;t)   \qquad\qquad\qquad\qquad \qquad\qquad \\
\nonumber
-\sum_{j}\int dx_{s+1}i L(x_j,x_{s+1})f_{s+1}(x_1\dots x_s,x_{s+1};t)   \qquad\qquad  \\
=iL_{s}\Delta f_{s}(x_1\dots x_s;t)-\sum_{j}\int dx_{s+1}
i L(x_j,x_{s+1})\Delta f_{s+1}(x_1\dots x_s,x_{s+1};t)  \nonumber    \\
+\int dx''\int da\int_{-\infty}^{t}dt'\mbox{e}^{\epsilon(t'-t)}
\phi_{G_{s}n}(x,x',x'',a;t,t')f(a;t')\gamma (x'';t')   \nonumber\\
\nonumber
-\sum_{\bf k}\int da\int_{-\infty}^{t}dt'\mbox{e}^{\epsilon (t'-t)}
\Big\{\phi_{G_{s}j}(x,x',{\bf k},a;t,t')\cdot \frac {\partial}{\partial{\bf J}_{\bf k}}  \qquad\quad \\
+  \,\,\phi_{G_{s}\varepsilon}(x,x',{\bf k},a;t,t')\frac{\partial}{\partial\varepsilon_{\bf k}}\Big\}
\frac{f(a;t')}{W(a;t')}\nonumber,    \qquad\qquad
\end{eqnarray}
where $\Delta f_{2}(x,x';t)=f_{2}(x,x';t)-g_{2}(x,x';t)$,
$\Delta f_{s}(x_1\dots x_{s};t)=f_{s}(x_1\dots x_{s};t)-g_{s}(x_1\dots x_{s};t)$.
In Eq.(\ref{Eq.30}) the two-particle Liouville operator
$$
iL_{2}=iL_{0}(x)+iL_{0}(x')+iL(x,x')
$$
was introduced. It contains one-particle operator
$$iL_{0}(x)=\frac{\bf p}{m}\cdot\frac{\partial}{\partial \bf r}, \quad x=\{{\bf r,p}\},$$
and also a potential part
$$
iL(x,x')=\frac{\partial }{\partial \bf r}\Phi(|{\bf r} -{\bf r}'|)
\cdot\Big[\frac{\partial }{\partial \bf p}-
\frac{\partial }{\partial {\bf p}'}\Big].
$$
Accordingly, in Eq.(\ref{Eq.31}), $iL_{s}$ is the $s$-particle Liouville operator:
$$
g_{s}(x_1\dots x_{s};t)=\langle \hat{G}_{s}(x_1\dots x_{s}) \rangle^{t}=
\int da g_{s}^{L}(x_1\dots x_{s};a;t)f(a;t),
$$
where
$$
g_{s}^{L}(x_1 \dots x_{s};a;t)=\int d\Gamma_{N}\hat{G}_{s}(x_1\dots x_{s})\varrho_{L}(x^{N};a;t)
$$
is the $s$-particle relevant distribution function of large-scale variables
and $\hat{G}_{s}(x^{s})=\hat{n}_{1}(x_{1})....\hat{n}_{1}(x_{s})$.

Thus we obtained  a system of equations for non-equilibrium
single, double, $s$-particle distribution functions
which take into account nonlinear hydrodynamic fluctuations.

We now discuss the equation (\ref{Eq.23}) that is of Fokker-Planck type one for non-equilibrium
distribution function of collective variables which take into account the kinetic processes.
The transport kernel in this equation $\phi_{nn}(x,x';t,t')$ describes a dissipation
of kinetic processes, while the kernels
$\phi_{nj}(x,{\bf k},a;t,t')$, $\phi_{n\varepsilon}(x,{\bf k},a;t,t')$, $\phi_{jn}(x,{\bf
k},a;t,t')$, $\phi_{\varepsilon n}(x,{\bf k},a;t,t')$ describe a dissipation
of correlations between kinetic and hydrodynamic processes.
To uncover  more detailed a structure  of transport
kernels $\phi _{nn}(x;x',a;t,t')$, $\phi _{Gn} (x;x',x'',a;t,t')$
we consider action of Liouville operator on
$\hat{n}_{1} (x)$ and  $\hat{G} (x,x')$  (see Appendix A).
That is, taking into account (\ref{Eq.23}) and (\ref{Eq.a5}), the kinetic equation (\ref{Eq.22})
can be written as follows:

\begin{eqnarray}    \label{Eq.38}
\Big[\frac{\partial}{\partial t}+\frac{\bf p}{m}
\frac{\partial}{\partial \bf r}\Big]
f_{1}(x;t)-\int dx'\int da \frac{\partial }{\partial \bf r}\Phi(|{\bf r} -{\bf r}'|) \qquad\qquad \\
\nonumber
\times\Big[\frac{\partial }{\partial \bf p}-
\frac{\partial }{\partial {\bf p}'}\Big]g_{2}^{l}(x,x',a;t)f(a;t)= \qquad\qquad\\
\nonumber
-\int dx'\int da\int_{-\infty}^{t}dt'e^{\epsilon(t'-t)}\frac{\partial }{\partial {\bf r}} \cdot D_{jj} (x,x',a;t,t')\cdot \frac{\partial }{\partial {\bf r}'}\gamma (x';t')f(a;t')       \\
\nonumber
+\int dx'\int da\int_{-\infty}^{t}dt'e^{\epsilon(t'-t)}\bigg[
\frac{\partial }{\partial {\bf p}} \cdot D_{Fj} (x,x',a;t,t')\cdot
\frac{\partial }{\partial {\bf r}'}        \qquad                    \\
\nonumber
+\frac{\partial }{\partial {\bf r}} \cdot D_{jF} (x,x',a;t,t')\cdot \frac{\partial }{\partial {\bf p}'} -\frac{\partial }{\partial {\bf p}} \cdot D_{FF} (x,x',a;t,t')\cdot \frac{\partial}{\partial {\bf p}'} \bigg]              \\
\nonumber
\times\gamma (x';t')f(a;t')-\sum_{\bf k}\int da\int_{-\infty}^{t}dt'e^{\epsilon (t'-t)}
\Big\{\phi_{nj}(x,{\bf k},a;t,t')\cdot \frac {\partial  }{\partial {\bf J}_{\bf k}}   \\
\nonumber
+\,\phi_{n\varepsilon}(x,{\bf k},a;t,t')\frac{\partial}{\partial\varepsilon_{\bf k}}\Big\}
\frac{f(a;t')}{W(a;t')}.  \qquad\qquad
\end{eqnarray}
In the equation (\ref{Eq.23}) the quantities $\phi_{jj}({\bf k},{\bf q},a,a';t,t')$,
$\phi_{j \varepsilon}({\bf k},{\bf q},a,a';t,t')$, $\phi_{\varepsilon j}({\bf k},{\bf
q},a,a';t,t')$, $\phi_{\varepsilon \varepsilon}({\bf k},{\bf q},a,a';t,t')$ correspond to
the dissipative processes connected with the correlations between viscous and heat
hydrodynamic processes.
The set of equations (\ref{Eq.22}), (\ref{Eq.23}), (\ref{Eq.30}), (\ref{Eq.31})
allows for two limiting cases.
First, if the description of non-equilibrium processes does not take into account
nonlinear hydrodynamic fluctuations, we will obtain generalized kinetic equation
for the non-equilibrium distribution function of the particles \cite{TokOmelKobr1998}:
\begin{eqnarray}       \label{Eq.33}
 \Big[\frac{\partial}{\partial t}+\frac{\bf p}{m}\cdot\frac{\partial}{\partial\bf r}\Big] f_{1}(x;t)-\int dx'\frac{\partial }{\partial \bf r}\Phi(|{\bf r} -{\bf r}'|)
 \cdot \Big[\frac{\partial }{\partial \bf p}-
 \frac{\partial }{\partial {\bf p}'}\Big]g_{2}(x,x';t)  \\
\nonumber
 = \int dx'\int\limits_{-\infty}^{t}dt' e^{\epsilon (t'-t)}\phi_{nn}(x,x';t,t')\gamma (x';t').\qquad\qquad
\end{eqnarray}
In this case, the Lagrange parameter $\gamma(x;t)$
is defined through the function $f_{1}(x;t)$ and,  essentially, the equation~(\ref{Eq.33})
is the closed kinetic equation for $f_{1}(x;t)$ with transport kernel
(memory function) $\Phi_{nn}(x,x';t,t')$, in which the Mori projection operator depends on time.
In the case of weakly non-equilibrium processes kinetic equation (\ref{Eq.33}) were obtained in
\cite{ForstMart1970,Mazenko1971} using Mori projection operator method.
Calculating function $\Phi_{nn}(x,x';t,t')$ by means of expansions over density,
the kinetic equations with linearized Boltzmann, Boltzmann-Enskog, Fokker-Planck
integrals were obtained. From this point of view these results can be related
with the results of BBGKY theory \cite{ZubMorRop1996:1}.
In our case the BBGKY chain of equations can be obtained assuming that
$\Delta f_{s}\approx 0$, namely $f_{s}$ close to $g_{s}$.

Second, if we do not take into account kinetic processes then we will obtain generalized
(non-Markov) Fokker-Planck equation for non-equilibrium distribution function
of collective variables, which can be obtained by the method of Zwanzig projection operators
or by the method of Zubarev non-equilibrium statistical operator \cite{ZubMor1983}:
\begin{eqnarray}    \label{Eq.34}
\frac{\partial }{\partial t}f(a;t)=\sum_{\bf k}\Big\{\frac {\partial }{\partial n_{{\bf k}}}v_{n,{\bf k}}(a;t)+
\frac {\partial  }{\partial {\bf J}_{\bf k}}\cdot {\bf v}_{j,{\bf k}}(a;t)+
\frac {\partial  }{\partial \varepsilon_{\bf k}}v_{\varepsilon,{\bf k}}(a;t)\Big\}f(a;t)    \\
\nonumber
=\sum_{\bf k,\bf q}\int da'\int\limits_{-\infty}^{t}dt' e^{\epsilon (t'-t)}
\frac {\partial  }{\partial {\bf J}_{\bf k}}\cdot \phi_{jj}({\bf k},{\bf q},a,a';t,t')
\cdot\frac {\partial  }{\partial {\bf J}_{\bf q}}\frac {f(a;t')}{W(a;t')}   \quad  \\
\nonumber
+\sum_{\bf k,\bf q}\int da'\int\limits_{-\infty}^{t}dt' e^{\epsilon (t'-t)}
\frac {\partial  }{\partial {\varepsilon}_{\bf k}}
\phi_{\varepsilon \varepsilon}({\bf k},{\bf q},a,a';t,t')
\frac {\partial  }{\partial {\varepsilon}_{\bf q}}\frac {f(a;t')}{W(a;t')} \qquad \\
\nonumber
+\sum_{\bf k,\bf q}\int da'\int\limits_{-\infty}^{t}dt' e^{\epsilon (t'-t)}
\Big\{\frac {\partial  }{\partial {\bf J}_{\bf k}}
 \cdot \phi_{j \varepsilon}({\bf k},{\bf q},a,a';t,t')
 \frac {\partial}{\partial {\varepsilon}_{\bf q}}  \qquad \\
\nonumber
 +\,\frac {\partial}{\partial {\varepsilon}_{\bf k}}
\phi_{\varepsilon j}({\bf k},{\bf q},a,a';t,t')
\cdot\frac{\partial}{\partial{\bf J}_{\bf q}}\Big\}
\frac {f(a;t')}{W(a;t')}.    \qquad\qquad
\end{eqnarray}

Now we give two examples of possible applications of the proposed approach   to study
the dynamics of simple liquids by taking into account nonlinear hydrodynamic fluctuations.
One of these problems is the consistent description of kinetic and hydrodynamic fluctuations
in turbulence phenomena in gases and liquids.
The method of non-equilibrium statistical operator for the description of turbulent flow
was proposed by Zubarev \cite{Zubarev1981,Zubarev1982}.
Main idea was to separate the system with the turbulent flow into two subsystems depending
on the scale and the character of the motions in the same space region. One of these subsystems,
that corresponds to large scale and regular movement, is described by hydrodynamic equations.
The second subsystem, that meets to smaller scale and movements with pulsations,
is described by statistical methods based on Fokker-Planck equation for
the distribution functional of the densities of particle number, momentum and energy.
The interaction between the subsystems leads to an exchange of momentum and energy between them.
The proposed here approach allows to describe self-consistently the kinetic and hydrodynamic
processes in calculations of the distribution functional of densities and $s$-particle
distribution functions from the equations (\ref{Eq.23}),(\ref{Eq.30}),(\ref{Eq.31}).
The need for consistent description of kinetic and hydrodynamic nonlinear
fluctuations in the investigation of turbulence phenomena in gases and liquids
were substantiated by Klimontovich \cite{Klimontov1996,Klimontov2002}.
The other statistical theory in which turbulence is considered as non-equilibrium phase
transition was proposed by Zubarev, Morozov and Troshkin \cite{ZubMorTrosh1992}.

Another interesting example might include using the proposed approach for consistent description
of kinetic and hydrodynamic processes to study the phase transitions in non-equilibrium
heterophase systems, in particular the liquid-gas (vapor) or liquid-glass systems.
The heterophase states of physical systems were investigated in many papers
\cite{BogShumYuk1984,ShumYuk1985,Yukalov1991,Fisher1993,OlemKopl1995,Olemskoi1999,
BakFish2004,Onuki2004,Das2011}.
We know that when heated a water to boiling temperature in it formed vapor bubbles.
On the other hand, above $0^{\circ}$C water is a mixture of water and ice,
which are present in macroscopic quantities but are not observed due to the rapid fluctuations in the system.
Namely, the presence of ice with structure different from the structure of water (liquid)
can explain the anomaly density of water in the temperature range from
$0^{\circ}$C  to $4^{\circ}$C.
They are the real heterophase systems in which bubble embryoses, drops or small crystals
have a kinetic nature caused by nonlinear fluctuations, changes in temperature, pressure.
In our case, the heterophase formations (containing a finite particle number in one or another
phase) can be described by non-equilibrium distribution function $f_{s}(x^{s};t)$.
The kinetic processes within heterophase formations are described respectively by the kinetic
equations (\ref{Eq.31}), in which right side contains the summands that take into account
the mutual influence of kinetic and hydrodynamic processes.
Obviously, such heterophase formations form and decay (with finite lifetime),
by exchanging with particles and energy with the surrounding particles on background
nonlinear hydrodynamic fluctuations of densities of particle number, momentum, energy,
the contribution of which increases at phase transformations.
These nonlinear fluctuations are  described by the Fokker-Planck equation (\ref{Eq.23}).
In the process of interaction between kinetic and hydrodynamic fluctuations
in heterophase systems with some particular change of temperature and pressure,
the self-organizing particle motions might occur due to spontaneous symmetry breaking.
These particle motions with group velocity
$f_{s}(x^{s};t)=f_{s}({\bf r}_{1}-{\bf v}t,{\bf p}_{1},\ldots,{\bf r}_{s}-{\bf v}t,{\bf p}_{s})$
might lead to an automodel (quasi-soliton) propagation of heterophase formations in the system.
Such processes require separate detailed study due to complex calculative
problems of kinetic and hydrodynamic transport kernels in transport equations.
In connection to these processes, we would like to draw attention to the
article by Klimontovich \cite{Klimontov1998},
in which to a certain extent a consistent description of kinetics
and hydrodynamics (taking account diffusion processes)
for gas-liquid phase transition is realized.
In the case in which we do not take into account the fluctuations of momentum
and energy in our equations, we will arrive for a similar Klimontovich equations.
These studies require separate consideration, in which the
structure function $W(a;t)$ of collective variables and of the hydrodynamic
velocities $v_{n,{\bf k}}(a;t)$, ${\bf v}_{j,{\bf k}}(a;t)$, $v_{\varepsilon,{\bf k}}(a;t)$
should be calculated. In next section we will perform these calculation.

\section{Calculation of structure function $W(a;t)$ and hydrodynamical velocities
         $v_{l,{\bf k}}(a;t)$}

In the Kawasaki theory \cite{Kawasaki1976} of non-linear fluctuations, the structure function is
approximated by a gaussian dependence on collective variables. In this case, as it can be seen,
the hydrodynamic velocities $v_{l,{\bf k}}(a;t)$, $l=n,j,\varepsilon$ are
the linear function of $a$.
Other approach for calculation of hydrodynamical velocities $v_{l,{\bf k}}(a;t)$
was proposed on the basis of local thermodynamics\cite{ZubMor1983}.
The resulting expressions  are valid obviously at low frequencies and
for small values of the wave vector, when the ratios of the local thermodynamics
are valid. Structure function $W(a;t)$ and hydrodynamical velocities $v_{l,{\bf k}}(a;t)$
in a case of study of hydrodynamic fluctuations were calculated in \cite{Zubarev1982,IdzIgnTok1996}
using the method of collective variables \cite{YukhnHolov1980}. The basic idea of this approach
is that the structure function $W(a;t)$ and hydrodynamic velocities $v_{l,{\bf k}}(a;t)$
can be calculated in approximations higher than Gaussian. Next, we apply the method
of collective variables \cite{Zubarev1982,YukhnHolov1980,IdzIgnTok1996}
for calculating the structure function $W(a;t)$ and hydrodynamic velocities $v_{l,\bf k}(a;t)$.

First, we calculate the structure function $ W (a; t) $ for collective variables.
To do this, we use the integral representation for $\delta$-functions:
\begin{equation}    \label{Eq.41}
\hat{f}(a)=\int d\omega \exp \Big\{ -i\pi \sum_{l,\bf k}
\omega_{l,\bf k}(\hat{a}_{l,\bf k}-a_{l,\bf k})\Big\},\quad
l=n,{\bf j},\varepsilon.
\end{equation}
Next, using a cumulant expansion \cite{IdzIgnTok1996} for $W(a;t)$ one obtains:
\begin{eqnarray}    \label{Eq.42}
W(a;t)=\int d\Gamma_{N}\varrho_{rel}^{kin} (x^{N};t)\hat{f}(a)   \qquad\qquad\qquad\qquad\qquad\\
\nonumber
=\int d\omega \exp \Big\{ -i\pi \sum_{l,\bf k}\omega_{l,\bf k}\bar{a}_{l,\bf k}-
\frac{\pi^{2}}{2}\sum_{l_{1},l_{2}}\sum_{{\bf k}_{1},{\bf k}_{2}}
\mathfrak{M}_{2}^{l_{1},l_{2}}({\bf k}_{1},{\bf k}_{2};t)\omega_{l_{1},{\bf k}_{1}}
 \omega_{l_{2},{\bf k}_{2}}\Big\}  \\
\nonumber
\times \exp\Big\{ \sum_{n\geq 3}D_{n}(\omega ;t)\Big\},  \qquad\qquad\qquad
\end{eqnarray}
where
\begin{equation}      \nonumber
\bar{a}_{l,\bf k}=a_{l,\bf k}-\langle\hat{a}_{l,\bf k}\rangle_{kin}^{t}, \,\,
 d\omega=\prod_{l,\bf k}d\omega_{l,\bf k}^{r}d\omega_{l,\bf k}^{s}, \,\,
 \omega_{l,\bf k}=\omega_{l,\bf k}^{r}-i\omega_{l,\bf k}^{s},\,\,
 \omega_{l,-\bf k}=\omega_{l,\bf k}^{*},
\end{equation}
\begin{equation}    \label{Eq.43}
 D_{n}(\omega ;t)=\frac{(-i\pi)^{n}}{n!}\sum_{l_{1},\ldots,l_{n}}\sum_{{\bf k}_{1},\ldots,{\bf k}_{n}}
 \mathfrak{M}_{n}^{l_{1},\ldots,l_{n}}({\bf k}_{1},\ldots,{\bf k}_{n};t)\omega_{l_{1},{\bf k}_{1}}\ldots.
 \omega_{l_{n},{\bf k}_{n}},
\end{equation}
\begin{equation}    \label{Eq.44}
\mathfrak{M}_{n}^{l_{1},\ldots,l_{n}}({\bf k}_{1},\ldots,{\bf k}_{n};t)=
\langle \hat{a}_{l_{1},{\bf k}_{1}},\ldots\hat{a}_{l_{n},{\bf k}_{n}}\rangle_{kin}^{t,c}
\end{equation}
are the non-equilibrium cumulant averages in approximations of the $n$-order, which are calculated
using distribution $\varrho_{rel}^{kin}(x^{N};t)$ (\ref{Eq.11}). We present the structure function
$W(a;t)$ for further calculations in following form:
\begin{eqnarray}    \label{Eq.45}
 W(a;t)=\int d\omega \exp \big\{-i\pi \sum_{l,\bf k}\omega_{l,\bf k}\bar{a}_{l,\bf k}  \\
\nonumber
 -\frac{\pi^{2}}{2}\sum_{l_{1},l_{2}}\sum_{{\bf k}_{1},{\bf k}_{2}}
 \mathfrak{M}_{2}^{l_{1},l_{2}}({\bf k}_{1},{\bf k}_{2};t)\omega_{l_{1},{\bf k}_{1}}
 \omega_{l_{2},{\bf k}_{2}}\big\}              \\
\nonumber
 \times\Big(1+B+\frac{1}{2!}B^{2}+\frac{1}{3!}B^{3}+\ldots+\frac{1}{n!}B^{n}+\dots\Big),
\end{eqnarray}
where $B=\sum_{n\geq 3}D_{n}(\omega ;t)$. If in series of exponent (\ref{Eq.45}), namely,
$\mbox{e}^{\sum_{n\geq 3}D_{n}(\omega;t)}$, one retains only the first term equal to unity,
one will obtain the Gaussian approximation for $W(a;t)$:
\begin{eqnarray}    \label{Eq.46}
  W^{G}(a;t)=\int d\omega \exp \{ i\pi \sum_{l,\bf k}\omega_{l,\bf k}\bar{a}_{l,\bf k}  \\
\nonumber
 - \frac{\pi^{2}}{2}\sum_{l_{1},l_{2}}\sum_{{\bf k}_{1},{\bf k}_{2}}
  \mathfrak{M}_{2}^{l_{1},l_{2}}({\bf k}_{1},{\bf k}_{2};t)
  \omega_{l_{1},{\bf k}_{1}} \omega_{l_{2},{\bf k}_{2}}\},
\end{eqnarray}
where $\mathfrak{M}_{2}^{l_{1},l_{2}}({\bf k}_{1},{\bf k}_{2};t)$ are the matrix elements of
non-equilibrium correlation functions:
\begin{equation}    \label{Eq.47}
\mathfrak{M}_{2}({\bf k}_{1},{\bf k}_{2};t)=
\begin{vmatrix}
 \langle \hat{n}\hat{n}\rangle_{kin}^{c}&
 \langle \hat{n}\hat{{\bf J}}\rangle_{kin}^{c} &
 \langle \hat{n}\hat{\varepsilon}\rangle_{kin}^{c}   \\
  \langle \hat{{\bf J}}\hat{n}\rangle_{kin}^{c}        &
  \langle \hat{{\bf J}}\hat{{\bf J}}\rangle_{kin}^{c}  &
  \langle \hat{{\bf J}}\hat{\varepsilon}\rangle_{kin}^{c}  \\
  \langle \hat{\varepsilon}\hat{n}\rangle_{kin}^{c}  &
  \langle \hat{\varepsilon}\hat{{\bf J}}\rangle_{kin}^{c}  &
  \langle \hat{\varepsilon}\hat{\varepsilon}\rangle_{kin}^{c}
\end{vmatrix}_{\lefteqn{\scriptstyle({\bf k_1,k_2})}},
\end{equation}
and the non-equilibrium cumulant average
\begin{equation}    \label{Eq.48}
\langle \hat{n}_{\bf k} \hat{n}_{-\bf k}\rangle_{kin}^{t,c}=
\langle \hat{n}_{\bf k} \hat{n}_{-\bf k}\rangle_{kin}^{t}-
\langle \hat{n}_{\bf k} \rangle_{kin}^{t}\langle  \hat{n}_{-\bf k}\rangle_{kin}^{t}.
\end{equation}
For integrating over $d\omega$ in (\ref{Eq.45}) we should transform the quadratic form
in exponential expression into a diagonal form with respect to $\omega_{l,{\bf k}}$.
To this end it is necessary to find the eigenvalues of the matrix (\ref{Eq.47}) by solving the equation
$$
\mbox{det} \big|\tilde{\mathfrak{M}}_{2}({\bf k}_{1},{\bf k}_{2};t)-\tilde{E}({\bf k}_{1},{\bf k}_{2};t)\big|=0,
$$
$\tilde{E}({\bf k}_{1},{\bf k}_{2};t)$ is the diagonal matrix.
Further, obtained eigenvalues $E_{l}({\bf k};t)$, $l=1, \ldots, 5$
of the expression (\ref{Eq.46}) are as follows:
\begin{eqnarray}    \label{Eq.49}
W^{G}(a;t)=\int d\tilde{\omega}\, \mbox{det}\tilde{W}\,
 \exp \Big\{ -i\pi \sum_{l,\bf k}   \tilde{a}_{l{\bf k}}
 \tilde{\omega}_{l{\bf k}}                                     \\
\nonumber
-\, \frac{\pi^{2}}{2}\sum_{l} \sum_{{\bf k}}
E_{l}({\bf k};t)\tilde{\omega}_{l{\bf k}} \tilde{\omega}_{l,-{\bf k}}\Big\},
\end{eqnarray}
where new variables $\tilde{a}_{l{\bf k}}$, $\tilde{\omega}_{l{\bf k}}$
are connected with the old variables by ratio:
$$ \tilde{a}_{n{\bf k}}=\sum_{l} \bar{a}_{l{\bf k}}\omega_{ln} , \qquad
   \omega_{l{\bf k}}=\sum^{3}_{m=1}\omega_{lm}\tilde{\omega}_{m{\bf k}},
$$
and $\omega_{lm}$ are matrix elements
$\tilde{W}=\begin{vmatrix}
 \omega_{11}&  \ldots & \omega_{15} \\
  \vdots    &  \ddots & \vdots      \\
 \omega_{51} & \ldots & \omega_{55}
\end{vmatrix}_{\lefteqn{\scriptstyle({\bf k};t)}}.$
Integrand in (\ref{Eq.49}) is a quadratic function $\tilde{\omega}_{n{\bf k}}$ and
after integrating over $d\omega_{n{\bf k}}$ we will obtain following structural function
in Gaussian approximation $W^{G}(a;t)$:
\begin{eqnarray}    \label{Eq.50}
 W^{G}(a;t)= \exp\big\{-\frac{1}{2}\sum_{l,\bf k} E^{-1}_{l}({\bf k};t)
  \tilde{a}_{l{\bf k}}\tilde{a}_{l,-{\bf k}}  \big\}     \qquad                 \\
\nonumber
\times\,  \exp\big\{-\frac{1}{2} \sum_{\bf k} \ln\,\pi \,\det \tilde{E}({\bf k};t) \big\}
  \exp \big\{\sum_{\bf k}\ln\,\det \tilde{W}({\bf k};t) \big\},
\end{eqnarray}
or through variables $\bar{a}_{l{\bf k}}$:
\begin{equation}    \label{Eq.51}
 W^{G}(a;t)= Z(t)\,\exp \big\{ -\frac{1}{2}\sum_{l,\bf k} \bar{E}_{l}({\bf k};t)
  \bar{a}_{l{\bf k}}\bar{a}_{l,-{\bf k}}   \big\},
\end{equation}
where
$$
\bar{E}_{l}({\bf k};t)=\sum_{l'}\omega_{ll'}E^{-1}_{l'}({\bf k};t)\omega_{l'l},
$$
$$
Z(t)= \exp \big\{-\frac{1}{2} \sum_{\bf k} \ln\,\pi \,\det \tilde{E}({\bf k};t) \big\}
  \exp \big\{\sum_{\bf k} \ln\,\det \tilde{W}({\bf k};t) \big\}.
$$

Gaussian approximation, corresponding to Kawasaki theory, is obtained as follows.
Kinetic processes are ignored in the calculations, then the average values
are calculated with microcanonical ensemble $ E = const $.
During the calculation
$\langle \hat{n}\hat{\bf J}  \rangle^c_E=0$,
$\langle \hat{\varepsilon}\hat{\bf J}  \rangle^c_E=0$
and the matrix has following form:
\begin{equation}    \nonumber
\mathfrak{M}_{2}({\bf k}_{1},{\bf k}_{2})=
\begin{vmatrix}
 \langle \hat{n}\hat{n}\rangle^c_E&0&\langle \hat{n}\hat{\varepsilon}\rangle^c_E\\
 0&\langle \hat{{\bf J}}\hat{{\bf J}}\rangle^{c}_E&0\\
 \langle \hat{\varepsilon}\hat{n}\rangle^{c}_E&0&
 \langle \hat{\varepsilon}\hat{\varepsilon}\rangle^{c}_E
\end{vmatrix}_{\lefteqn{\scriptstyle({\bf k_1,k_2})}}.
\end{equation}
In this case a diagonalization of quadratic form in $W^G(a;t)$ occurs at the transition
from the variables $\hat{n},\,\hat{\bf J},\,\hat{\varepsilon}$ to
$\hat{n},\,\hat{\bf J},\,\hat{h}$
\begin{equation}    \nonumber
 \hat{h}_{\bf k}= \hat{\varepsilon}_{\bf k} -
   \frac{\langle\hat{\varepsilon}_{\bf k}\hat{n}_{\bf -k}\rangle^c_E}
      {\langle\hat{n}_{\bf k}\hat{n}_{\bf -k}\rangle^c_E}\, \hat{n}_{\bf k},
\end{equation}
where $\hat{h}_{\bf k}$ is the Fourier component of generalized enthalpy.
After these transformations Gaussian dependence is obtained as
 $W^G(a;t)\sim \exp\{-\frac{\hat{a}^2}{\langle \hat{a}^2 \rangle}\}.$

The structure function $W^{G}(a;t)$ gives a possibility to calculate
(\ref{Eq.45}) in higher approximations over Gaussian moments \cite{IdzIgnTok1996}:
\begin{equation}    \label{Eq.52}
 W(a;t)=W^{G}(a;t)\exp \big\{\sum_{n\geq 3}\langle \tilde{D}_{n}(a ;t)\rangle_{G}\big\},
\end{equation}
where one presents $\langle\tilde{D}_{n}(a;t)\rangle_{G}$ approximately as:
$$
\langle\tilde{D}_{3}(a ;t)\rangle_{G}=\langle \bar{D}_{3}(a ;t)\rangle_{G},
$$
$$
\langle\tilde{D}_{4}(a ;t)\rangle_{G}=\langle \bar{D}_{4}(a ;t)\rangle_{G},
$$
$$
\langle\tilde{D}_{6}(a ;t)\rangle_{G}=\langle \bar{D}_{6}(a ;t)\rangle_{G}-
\frac{1}{2}\langle\bar{D}_{3}(a ;t)\rangle_{G}^{2},
$$
$$
\langle\tilde{D}_{8}(a ;t)\rangle_{G}=\langle \bar{D}_{8}(a ;t)\rangle_{G}-
\langle\bar{D}_{3}(a;t)\rangle_{G}\langle \bar{D}_{5}(a ;t)\rangle_{G} -
\frac{1}{2}\langle \bar{D}_{4}(a ;t)\rangle_{G}^{2},
$$
$$
\langle\tilde{D}_{n}(a;t)\rangle_{G}=
\frac{1}{W^{G}(a;t)}\sum_{l_{1},\ldots,l_{n}}\sum_{{\bf k}_{1},\ldots,{\bf k}_{n}}
\bar{\mathfrak{M}}_{n}^{l_{1},\ldots,l_{n}}({\bf k}_{1},\ldots,{\bf k}_{n};t)
$$
$$
\times\, \frac{1}{(i\pi)^{n}}\frac{\delta^{n}}{\delta \bar{a}_{l_{1},{\bf k}_{1}}...
\delta \bar{a}_{l_{n},{\bf k}_{n}}}W^{G}(a;t).
$$

\noindent
$\bar{\mathfrak{M}}_{n}^{l_{1},\ldots,l_{n}}({\bf k}_{1},\ldots,{\bf k}_{n};t)$ are the
renormalized non-equilibrium cumulant averages of order $n$ for the variables
$\bar{a}_ {l{\bf k}}$. In expression (\ref{Eq.52}) the summands
are with only even degrees over ${a}$ since all odd Gaussian moments vanish.

The method of calculation of the structure function $W(a;t)$ can be applied for approximate
calculations of hydrodynamic velocities $v_{l,{\bf k}}(a;t)$. We present general formula
of velocities consistent with (\ref{Eq.29}) in following form:
$$
v_{l,{\bf k}}(a;t)=\int d\Gamma_{N}\dot{\hat{a}}_{l,{\bf k}}\varrho_{rel}^{kin} (x^{N};t)\hat{f}(a)
$$
and introduce function $W(a,\lambda;t)$:
$$
W(a,\lambda;t)=\int d\Gamma_{N}\,\,\mbox{e}^{-i\pi\sum_{l,\bf k}\lambda_{l,\bf k}\dot{\hat{a}}_{l,{\bf k}}}\,\,\varrho_{rel}^{kin} (x^{N};t)\hat{f}(a),
$$
so that
\begin{equation}    \label{Eq.53}
v_{l,{\bf k}}(a;t)=\frac{\partial }{\partial (-i\pi\lambda_{l,\bf k})}
\ln W(a,\lambda;t)\Big|_{\lefteqn{\scriptstyle\lambda_{l,\bf k}=0}}.
\end{equation}
We calculate the function $W(a,\lambda;t)$ using the preliminary results
of the calculation of the structural function $W(a;t)$,
and rewrite $W(a,\lambda;t)$ as:
\begin{eqnarray}    \label{Eq.54}
 W(a,\lambda;t)=\int d\Gamma_{N}\int d\omega
 \exp \Big\{-i\pi\sum_{l,\bf k}\lambda_{l,\bf k}\dot{\hat{a}}_{l,{\bf k}}\Big\}  \\
\nonumber
\times\, \exp \Big\{ -i\pi \sum_{l,\bf k}\omega_{l,\bf k}(\hat{a}_{l,\bf k}-a_{l,\bf k})\Big\}
 \varrho_{rel}^{kin} (x^{N};t).
\end{eqnarray}
Now we carry out an averaging in (\ref{Eq.54}) with $\varrho_{rel}^{kin}(x^{N};t)$
using following cumulant expansion:
\begin{eqnarray}    \label{Eq.55}
W(a,\lambda;t)=
\int d\omega \exp \Big\{ -i\pi \sum_{l,\bf k}\omega_{l,\bf k}\bar{a}_{l,\bf k}    \\
\nonumber
 +\, \sum_{n\geq 1}\left[D_{n}(\omega ;t)+D_{n}(\lambda ;t)+D_{n}(\omega,\lambda ;t)\right]\Big\},
\end{eqnarray}
where
\begin{eqnarray}         \nonumber
 D_{n}(\omega ;t)=\frac{(-i\pi)^{n}}{n!}\sum_{l_{1},\ldots,l_{n}}\sum_{{\bf k}_{1},\ldots,
   {\bf{k}}_{n}}   \mathfrak{M}_{n}^{l_{1},\ldots,l_{n}}({\bf k}_{1},\ldots,{\bf k}_{n};t)
   \omega_{l_{1},{\bf k}_{1}}\ldots \omega_{l_{n},{\bf k}_{n}},       \\
\nonumber
D_{n}(\lambda ;t)=\frac{(-i\pi)^{n}}{n!}\sum_{l_{1},\ldots,l_{n}}\sum_{{\bf k}_{1},\ldots,
{\bf k}_{n}} \mathfrak{M}_{n}^{(1)l_{1},\ldots,l_{n}}({\bf k}_{1},\ldots,
{\bf k}_{n};t)\lambda_{l_{1},{\bf k}_{1}}\ldots\lambda_{l_{n},{\bf k}_{n}},     \\
\nonumber
D_{n}(\omega, \lambda ;t)=\frac{(-i\pi)^{n}}{n!}
\sum_{l_{1},\ldots,l_{n}}\sum_{{\bf k}_{1},\ldots,{\bf k}_{n}}
 \mathfrak{M}_{n}^{(2)l_{1},\ldots,l_{n}}({\bf k}_{1},\ldots,{\bf k}_{n};t)   \qquad       \\
\nonumber
\times\, \omega_{l_{1},{\bf k}_{1}}\ldots \omega_{l_{n-1},{\bf k}_{n-1}}\ldots
\lambda_{l_{n},{\bf k}_{n}},              \qquad\qquad
\end{eqnarray}
with the cumulants of following structure:
\begin{eqnarray}
\nonumber
\mathfrak{M}_{n}^{l_{1},\ldots,l_{n}}({\bf k}_{1},\ldots,{\bf k}_{n};t)=
\langle \hat{a}_{l_{1},{\bf k}_{1}},\ldots\hat{a}_{l_{n},{\bf k}_{n}}\rangle_{kin}^{t,c},  \\
\nonumber
 \mathfrak{M}_{n}^{(1)l_{1},\ldots,l_{n}}({\bf k}_{1},\ldots,{\bf k}_{n};t)=
 \langle\dot{\hat{a}}_{l_{1},{\bf k}_{1}},\ldots\dot{\hat{a}}_{l_{n},{\bf k}_{n}}\rangle_{kin}^{t,c},                                                   \\
\nonumber
 \mathfrak{M}_{n}^{(2)l_{1},\ldots,l_{n}}({\bf k}_{1},\ldots,{\bf k}_{n};t)=n[(n-j)+(j-n+1)
 \delta_{l_{1},\ldots,l_{n-1}}]              \\
\nonumber
\times\, \langle \hat{a}_{l_{1},{\bf k}_{1}},\ldots,\hat{a}_{l_{n-j},{\bf k}_{n-j}},\ldots,
 \dot{\hat{a}}_{l_{n-j+1},{\bf k}_{n-j+1}},\ldots,\dot{\hat{a}}_{l_{n},{\bf k}_{n}}\rangle_{kin}^{t,c}.
\end{eqnarray}
First, we consider a Gaussian approximation for $W(a,\lambda;t)$,
namely we leave in the exponent of
an integrand only the summands with $n=2$ and linear over $\lambda_{l,{\bf k}}$:
\begin{eqnarray}    \label{Eq.56}
  W^{G}(a,\lambda;t)=\int d\omega\exp \Big\{ i\pi \sum_{l,{\bf k}}
  \omega_{l,{\bf k}}\bar{a}_{l,\bf k}-
i\pi \sum_{l,\bf k}\langle\dot{\hat{a}}_{l,{\bf k}}\rangle_{kin}^{t,c}\lambda_{l,{\bf k}}    \\
\nonumber
  -\,\frac{\pi^{2}}{2}\sum_{l_{1},l_{2}}\sum_{{\bf k}_{1},{\bf k}_{2}}
  \mathfrak{M}_{2}^{l_{1},l_{2}}({\bf k}_{1},{\bf k}_{2};t)
  \omega_{l_{1},{\bf k}_{1}} \omega_{l_{2},{\bf k}_{2}}                 \\
\nonumber
-\,\frac{\pi^{2}}{2}\sum_{l_{1},l_{2}}\sum_{{\bf k}_{1},{\bf k}_{2}}
  \mathfrak{M}_{2}^{(2)l_{1},l_{2}}({\bf k}_{1},{\bf k}_{2};t)
  \omega_{l_{1},{\bf k}_{1}} \lambda_{l_{2},{\bf k}_{2}}\Big\}.
\end{eqnarray}
Then, transforming this expression in the exponent to diagonal quadratic form over variables
$\omega_{l,{\bf k}}$, similarly as for $W(a;t)$, after integrating with respect to the new variables
$\bar{\omega}_{l,{\bf k}}$, one obtains:
\begin{eqnarray}    \label{Eq.57}
 W^{G}(a,\lambda;t)=\int d\omega \exp \Big\{-i\pi \sum_{l,{\bf k}}
  \langle\dot{\hat{a}}_{l,{\bf k}}\rangle_{kin}^{t}\lambda_{l,{\bf k}}-
 \frac{\pi^{2}}{2}\sum_{l,{\bf k}}E_{l}^{-1}({\bf k};t)b_{l,{\bf k}}b_{l,-{\bf k}}    \\
\nonumber
-\,\frac{1}{2}\sum_{{\bf k}}\ln\pi det\tilde{E}({\bf k};t)+\sum_{{\bf k}}\ln det\tilde{W}({\bf k};t)\Big\},   \qquad\qquad
\end{eqnarray}
where
$$
b_{l,{\bf k}}=\sum_{j}\omega_{lj}\Big[\bar{a}_{j,{\bf k}}+
\frac{i\pi}{2}\sum_{j'}\mathfrak{M}_{2}^{(2)j,j'}({\bf k};t)\lambda_{j',{\bf k}}\Big],
$$
and $\omega_{lj}$, $\mathfrak{M}_{2}^{(2)j,j'}({\bf k};t)$ and $E_{l}({\bf k};t)$ do not
dependent on $\lambda_{l,{\bf k}}$. Here the cumulant  $\mathfrak{M}_{2}^{(2)j,j'}({\bf k};t)$
has the following structure:
\begin{equation}    \label{Eq.58}
\mathfrak{M}_{2}^{(2)j,j'}({\bf k};t)=
    \langle\dot{\hat{a}}_{j,{\bf k}}\hat{a}_{j',-{\bf k}}\rangle_{kin}^{t}-
    \langle\dot{\hat{a}}_{j,{\bf k}}\rangle_{kin}^{t}\langle\hat{a}_{j',-{\bf k}}\rangle_{kin}^{t}.
\end{equation}
Now we calculate the hydrodynamic velocities $v_{l,{\bf k}}(a;t)$ in Gaussian approximation
according to the formula (\ref{Eq.53}) :
\begin{eqnarray}    \label{Eq.59}
v_{l,{\bf k}}(a;t)=\frac{\partial }{\partial (-i\pi\lambda_{l,\bf k})}
\ln W^{G}(a,\lambda;t)\Big|_{\lambda_{l,{\bf k}}=0}                   \\
\nonumber
=\,\langle\dot{\hat{a}}_{j,{\bf k}}\rangle_{kin}^{t}-
\frac{1}{2}\sum_{j,j'}E_{l}^{-1}({\bf k};t)\omega_{jl}
\omega_{j'l}\mathfrak{M}_{2}^{(2)j',l}({\bf k};t)\bar{a}_{l,{\bf k}}.
\end{eqnarray}
Specifically, we consider the particular case when one can divide the longitudinal and transverse fluctuations for collective variables. That is, we choose the direction of the wave vector ${\bf k}$
along the axis of $0z$. Thus, one obtains:
\begin{equation}    \label{Eq.60}
  W^{G}(a;t)=\int d\omega \exp \{ i\pi \sum_{l,\bf k}\omega_{l,\bf k}\bar{a}_{l,\bf k}-
  \frac{\pi^{2}}{2}\sum_{l_{1},l_{2}=1}^{3}\sum_{{\bf k}_{1},{\bf k}_{2}}
  \mathfrak{M}^{\parallel, l_{1},l_{2}}_{2}({\bf k}_{1},{\bf k}_{2};t)
  \omega_{l_{1},{\bf k}_{1}} \omega_{l_{2},{\bf k}_{2}}
\end{equation}
$$
-\frac{\pi^{2}}{2}\sum_{l_{1},l_{2}=1}^{4}\sum_{{\bf k}_{1},{\bf k}_{2}}
  \mathfrak{M}^{\parallel,\perp, l_{1},l_{2}}_{2}({\bf k}_{1},{\bf k}_{2};t)
  \omega_{l_{1},{\bf k}_{1}} \omega_{l_{2},{\bf k}_{2}}\},
$$
where $\mathfrak{M}^{\parallel,l_{1},l_{2}}_{2}({\bf k}_{1},{\bf k}_{2};t)$
are the matrix elements of the non-equilibrium correlation functions of longitudinal fluctuations
\begin{equation}    \label{Eq.61}
\mathfrak{M}^{\parallel}_{2}({\bf k}_{1},{\bf k}_{2};t)=
\begin{vmatrix}
 \langle \hat{n}\hat{n}\rangle_{kin}^{c}&
 \langle \hat{n}\hat{{\bf J}}^{\parallel}\rangle_{kin}^{c} &
 \langle \hat{n}\hat{\varepsilon}\rangle_{kin}^{c}   \\
  \langle \hat{{\bf J}}^{\parallel}\hat{n}\rangle_{kin}^{c}        &
  \langle \hat{{\bf J}}^{\parallel}\hat{{\bf J}}^{\parallel}\rangle_{kin}^{c}  &
  \langle \hat{{\bf J}}^{\parallel}\hat{\varepsilon}\rangle_{kin}^{c}  \\
  \langle \hat{\varepsilon}\hat{n}\rangle_{kin}^{c}  &
  \langle \hat{\varepsilon}\hat{{\bf J}}^{\parallel}\rangle_{kin}^{c}  &
  \langle \hat{\varepsilon}\hat{\varepsilon}\rangle_{kin}^{c}
  \end{vmatrix}_{\lefteqn{\scriptstyle({\bf k_1,k_2})}},
\end{equation}
and $\mathfrak{M}^{\perp l_{1},l_{2}}_{2}({\bf k}_{1},{\bf k}_{2};t)$
are the matrix elements of the non-equilibrium correlation functions of transverse and
transverse-longitudinal fluctuations
\begin{equation}    \label{Eq.62}
\mathfrak{M}^{\parallel,\perp}_{2}({\bf k}_{1},{\bf k}_{2};t)=
\begin{vmatrix}
 0&
 \langle \hat{n}\hat{{\bf J}}^{\perp}_{x}\rangle_{kin}^{c} &
 \langle \hat{n}\hat{{\bf J}}^{\perp}_{y}\rangle_{kin}^{c} &
 0   \\
 \langle \hat{{\bf J}}^{\perp}_{x}\hat{n}\rangle_{kin}^{c}        &
  \langle \hat{{\bf J}}^{\perp}_{x}\hat{{\bf J}}^{\perp}_{x}\rangle_{kin}^{c}  &
  \langle \hat{{\bf J}}^{\perp}_{x}\hat{{\bf J}}^{\perp}_{y}\rangle_{kin}^{c}  &
  \langle \hat{{\bf J}}^{\perp}_{x}\hat{\varepsilon}\rangle_{kin}^{c}  \\
  \langle \hat{{\bf J}}^{\perp}_{y}\hat{n}\rangle_{kin}^{c}        &
  \langle \hat{{\bf J}}^{\perp}_{y}\hat{{\bf J}}^{\perp}_{x}\rangle_{kin}^{c}  &
  \langle \hat{{\bf J}}^{\perp}_{y}\hat{{\bf J}}^{\perp}_{y}\rangle_{kin}^{c}  &
  \langle \hat{{\bf J}}^{\perp}_{y}\hat{\varepsilon}\rangle_{kin}^{c}  \\
  0  &
  \langle \hat{\varepsilon}\hat{{\bf J}}^{\perp}_{x}\rangle_{kin}^{c}  &
  \langle \hat{\varepsilon}\hat{{\bf J}}^{\perp}_{y}\rangle_{kin}^{c}  &
  0
\end{vmatrix}_{\lefteqn{\scriptstyle({\bf k_1,k_2})}}.
\end{equation}
In this case, the hydrodynamic velocities in the Gaussian approximation are as follows:
\begin{eqnarray}    \label{Eq.63}
\nonumber
 v_{n{\bf k}}^{\parallel G}(a;t)=\langle\dot{\hat{n}}_{{\bf k}}\rangle_{kin}^{t}+ E_{1}^{-1}({\bf
k};t)(\omega_{11}\bar{n}_{{\bf k}}+ \omega_{21}\bar{{\bf J}}^{\parallel}_{{\bf k}} + \omega_{31}
\bar{\varepsilon}_{{\bf k}})\Omega_{n}({\bf k};t),    \\
v_{J{\bf k}}^{\parallel G}(a;t)=\langle\dot{\hat{{\bf J}}}^{\parallel }_{{\bf k}}\rangle_{kin}^{t}+
E_{2}^{-1}({\bf k};t)(\omega_{12}\bar{n}_{{\bf k}}+\omega_{22}\bar{{\bf J}}^{\parallel }_{{\bf k}} +
\omega_{32} \bar{\varepsilon}_{{\bf k}})\Omega_{J}({\bf k};t),  \\
\nonumber
v_{\varepsilon{\bf k}}^{\parallel G}(a;t)=\langle\dot{\hat{\varepsilon}}_{{\bf k}}\rangle_{kin}^{t}+
E_{3}^{-1}({\bf k};t)(\omega_{13}\bar{n}_{{\bf k}}+ \omega_{23}\bar{{\bf J}}^{\parallel}_{{\bf k}} +
\omega_{33} \bar{\varepsilon}_{{\bf k}})\Omega_{\varepsilon}({\bf k};t),
\end{eqnarray}
where
\begin{eqnarray}     \label{Eq.64}
\nonumber
\Omega_{n}({\bf k};t)=\omega_{11}\langle \hat{n}_{{\bf k}}\dot{\hat{n}}_{-{\bf k}}
\rangle_{kin}^{t,c}+ \omega_{21}\langle \hat{{\bf J}}^{\parallel}_{{\bf k}}\dot{\hat{n}}_{-{\bf k}}
\rangle_{kin}^{t,c}+ \omega_{31}\langle \hat{\varepsilon}_{{\bf k}}\dot{\hat{n}}_{-{\bf k}}
\rangle_{kin}^{t,c},                    \\
\Omega_{J}({\bf k};t)=\omega_{12} \langle \hat{n}_{{\bf k}}\dot{\hat{{\bf J}}}^{\parallel}_{-{\bf k}}
\rangle_{kin}^{t,c}+\omega_{22}\langle\hat{{\bf J}}^{\parallel}_{{\bf k}}\dot{\hat{{\bf J}}}^{\parallel}_{-{\bf k}}
\rangle_{kin}^{t,c}+\omega_{32}\langle\hat{\varepsilon}_{{\bf k}} \dot{\hat{{\bf J}}}^{\parallel}_{-{\bf
k}} \rangle_{kin}^{t,c},                  \\
\nonumber
 \Omega_{\varepsilon}({\bf k};t)=\omega_{13} \langle \hat{n}_{{\bf k}}
 \dot{\hat{\varepsilon}}_{-{\bf k}}\rangle_{kin}^{t,c}+
 \omega_{23} \langle \hat{{\bf J}}^{\parallel}_{{\bf k}}
 \dot{\hat{\varepsilon}}_{-{\bf k}}\rangle_{kin}^{t,c}+
 \omega_{33}\langle\hat{\varepsilon}_{{\bf k}}
 \dot{\hat{\varepsilon}}_{-{\bf k}} \rangle_{kin}^{t,c},
\end{eqnarray}
and $\omega_{lj}$ are the elements of matrix $\tilde{W}({\bf k};t)$.
As one can be see, the hydrodynamic velocities (\ref{Eq.63})
in the Gaussian approximation for
$W^{G}(a,\lambda;t)$ are the linear functions of collective variables $n_{\bf k}$,
${\bf J}_{\bf k}$ and $\varepsilon_{\bf k}$.
It is remarkable that if the kinetic processes are not taken into account,
then $\varrho_{rel}^{kin} (x^{N};t)=1$ $\langle\ldots\rangle_{kin}^{t}\rightarrow\langle\ldots\rangle_{0}$
is an averaging over a microscopic ensemble $W(a)$;
in this case the expressions (\ref{Eq.63}) for hydrodynamic velocities
transform into the results of previous work \cite{IdzIgnTok1996},
in which the nonlinear hydrodynamic fluctuations in simple fluids were investigated.
The collective variable method \cite{Zubarev1982,YukhnHolov1980,IdzIgnTok1996}
give a possibility to calculate the hydrodynamic velocities in approximations higher
than the Gaussian one.
In particular, the approximation for the Gaussian, based on (\ref{Eq.55})
and hydrodynamic velocities (\ref{Eq.63}) will be proportional to
$\bar{a}_{l,{\bf k}}\bar{a}_{l',{\bf k}}$, and transport kernels
in the Fokker-Planck equation will be the fourth-order correlation
functions over the variables $\hat{a}_{l,{\bf k}}$.

It is important that in Gaussian approximation for $\tilde{W}^{G}({\bf k};t)$ and
$v_{l,{\bf k}}^{G}(a;t)$, the Fokker-Planck equation leads to the transport equations
for $\langle \hat{a}_{l,{\bf k}}\rangle^{t}$, which are similar in structure
to the case of the molecular hydrodynamics,
averaged only over $\varrho_{L}(x^{N},a;t)=\varrho_{rel}^{kin}(x^{N};t)\frac{\hat{f}(a)}{W^{G}(a;t)}$.
The proposed approach makes possible to go beyond the Gaussian approximation for
$\tilde{W}({\bf k};t) $ and $v_{l,{\bf k}} (a;t) $, and hence to do the same
in the transport kernels in Foker-Planck equation.
This allows us to obtain a nonlinear equation system for $\langle\hat{a}_{l,{\bf k}}\rangle^{t}$ .

It is noteworthy that kinetic equation (\ref{Eq.23}) contains a generalized integral
of Fokker-Planck type  with generalized coefficients of diffusion and
particle friction in the phase space $({\bf r},{\bf p},t)$.
This region of changes $|{\bf r}|$ is limited by values $|{\bf k}|^{-1}_{hydr}$,
that correspond to collective nonlinear hydrodynamic processes.
This means that in regions of limited $|{\bf k}|^{-1}_{hydr}$ the processes
are described by the generalized coefficients of diffusion and friction,
and at small $|{\bf k}|^{-1}_{hydr}$ they are described by generalized viscosity,
thermal conductivity and by cross coefficients
$\phi_{jj}({\bf k},{\bf q},a,a';t,t')$, $\phi_{j \varepsilon}({\bf k},{\bf q},a,a';t,t')$, $\phi_{\varepsilon j}({\bf k},{\bf q},a,a';t,t')$, $\phi_{\varepsilon \varepsilon}({\bf k},{\bf q},a,a';t,t')$. Correlations between these regions are described by cross kernels $\phi_{n j}(x,{\bf q},a,a';t,t')$, $\phi_{n \varepsilon}(x,{\bf q},a,a';t,t')$, $\phi_{ \varepsilon n}({\bf k},x',a,a';t,t')$, $\phi_{jn}({\bf k},x',$ $a,a';t,t')$, that are present both in the kinetic equation and in the Fokker-Planck equation.
The calculations of these kernels is very important because they describe the cross-correlations between kinetic and hydrodynamic processes.

\section{Conclusions}

Using the method of Zubarev non-equilibrium statistical operator,
we have developed an approach \cite{MorKobrTok1994,ZubMor1983}
for consistent description of kinetic and hydrodynamic processes,
that are characterized by non-linear fluctuations.
We have obtained the non-equilibrium statistical operator of non-equilibrium
state of the system when the parameters of the reduced description are a
non-equilibrium one-particle distribution function and the
non-equilibrium distribution function of the non-linear hydrodynamic variables
(densities of mass, momentum and energy).
By using this operator we constructed a chain of kinetic equations (of BBGKY type)
for non-equilibrium single, double, $s$-particle distribution functions of particles
that take into account the nonlinear hydrodynamic fluctuations.
At the same time the non-equilibrium distribution function
of hydrodynamic fluctuations satisfy a generalized Fokker-Planck equation.

We proposed a method to calculate the structural distribution function
of hydrodynamic collective variables and their hydrodynamic velocities
(above Gaussian approximation) contained in a generalized Fokker-Planck
equation for the non-equilibrium distribution function of hydrodynamic
collective variables. In the future studies, we will go beyond the Gaussian
approximation and carry out approximate calculations of kinetic transport
coefficients for a specific system of interacting particles.

The proposed approach of the consistent description of kinetic and hydrodynamic processes
can be applied for a description of turbulence phenomena in dense gases and complex liquids
when a kinetics of a certain component, such as behavior of neutrons in the scattering processes,
must be take into account.
It is appealing to apply this approach for description of kinetic and hydrodynamic
nonlinear fluctuations in dense dusty plasmas \cite{MarkTok2014}, in which an important problem
is to take into account the non-equilibrium electromagnetic processes.
We plan to use the outlined approach to investigate the non-equilibrium
processes in the liquid-vapor phase transition that is characterized by
non-equilibrium hydrodynamical fluctuations of masses, momentum and energy
of particles. The fluctuations describe the collective nature of processes
and define the time-spatial behavior of  transport coefficients
of viscosity, thermal conductivity and the process of the emergence of liquid drops
in the gas phase or the gas bubbles in the liquid phase that has kinetic
nature.

\appendix
\section{}

Now we consider action of Liouville operator on
$\hat{n}_{1} (x)$ and  $\hat{G} (x,x')$:

\begin{eqnarray}            \label{Eq.a1}
iL_{N} \hat{n}_{1} (x)=-\frac{\partial }{\partial {\bf r}} \cdot \frac{1}{m} \hat{{\bf j}}({\bf r},{\bf p})+\frac{\partial }{\partial {\bf p}} \cdot \hat{{\bf F}}({\bf r},{\bf p}),
\end{eqnarray}
\begin{eqnarray}            \label{Eq.a2}
iL_{N} \hat{G} (x,x')&=&-\frac{\partial }{\partial {\bf r}} \cdot \frac{1}{m} \hat{{\bf j}}({\bf r},{\bf p})\hat{n}_{1} (x')-\hat{n}_{1} (x)\frac{\partial }{\partial {\bf r}'} \cdot \frac{1}{m} \hat{{\bf j}}({\bf r}',{\bf p}') \\
&&+\frac{\partial }{\partial {\bf p}} \cdot \hat{{\bf F}}({\bf r},{\bf p})\hat{n}_{1} (x')+\hat{n}_{1} (x)\frac{\partial }{\partial {\bf p}'} \cdot \hat{{\bf F}}({\bf r}',{\bf p}'),\nonumber
\end{eqnarray}
where
\begin{equation}           \label{Eq.a3}
\hat{{\bf j}}({\bf r},{\bf p})=\sum _{j=1}^{N} {\bf p}_{j} \delta ({\bf r}-{\bf r}_{j} )\delta ({\bf p}-{\bf p}_{j} )
\end{equation}
is the microscopic density of momentum vector in coordinate-momentum space,
\begin{equation}            \label{Eq.a4}
\hat{{\bf F}}({\bf r},{\bf p})=\sum _{l\ne j} \frac{\partial }{\partial {\bf r}_{j} } \Phi (|{\bf r}_{j} -{\bf r}_{l} |)\delta ({\bf r}-{\bf r}_{j} )\delta ({\bf p}-{\bf p}_{j} )
\end{equation}
is the microscopic density of force vector in coordinate-momentum space.

Taking into account equations (\ref{Eq.a1})-(\ref{Eq.a4}),
for the kinetic transport kernels, we obtain:
\begin{eqnarray}             \label{Eq.a5}
\phi _{nn} (x;x',a;t,t')=
-\bigg[\frac{\partial}{\partial{\bf r}}\cdot D_{jj} (x,x',a;t,t')\cdot\frac{\partial }{\partial {\bf r}'}       \qquad\qquad        \\
- \frac{\partial}{\partial{\bf p}}\cdot D_{Fj}(x,x',a;t,t')\cdot\frac{\partial}{\partial{\bf r}'}  \nonumber         \qquad\qquad\qquad\qquad  \\
-\frac{\partial }{\partial {\bf r}} \cdot D_{jF} (x,x',a;t,t')\cdot \frac{\partial }{\partial {\bf p}'} +\frac{\partial }{\partial {\bf p}} \cdot D_{FF} (x,x',a;t,t')\cdot \frac{\partial }{\partial {\bf p}'} \bigg],  \nonumber
\end{eqnarray}
where
\begin{equation}         \nonumber
D_{jj} (x,x',a;t,t')=\int d\Gamma _{N} \hat{{\bf j}}(x)T(t,t')(1-P(t'))\hat{{\bf j}}(x')\rho _{L} (x^{N} ;t'),
\end{equation}
\begin{equation}         \nonumber
D_{FF} (x,x',a;t,t')=\int d\Gamma _{N} \hat{{\bf F}}(x)T(t,t')(1-P(t'))\hat{{\bf F}}(x')\rho _{L} (x^{N} ;t')
\end{equation}
are the generalized diffusion and the particle friction coefficients
in the coordinate-momentum space. Moreover,
\begin{equation}         \nonumber
\int d{\bf p}\int d{\bf p}'  D_{jj} (x,x';t,t')=D_{jj} ({\bf r},{\bf r}';t,t'),
\end{equation}
\begin{equation}          \nonumber
\int d{\bf p}\int d{\bf p}'  D_{FF} (x,x';t,t')=D_{FF} ({\bf r},{\bf r}';t,t')
\end{equation}
are the generalized coefficients of diffusion and friction. Similarly, we obtain the expression for the transport kernel $\phi_{Gn}(x;x',x'';t,t')$:

 \begin{eqnarray}          \label{Eq.a6}
\phi _{Gn} (x;x',x'',a;t,t')=
-\bigg[\frac{\partial }{\partial {\bf r}} \cdot D_{jjn} (x,x',x'',a;t,t')\cdot \frac{\partial}{\partial {\bf r}'}  \qquad\qquad\\
+\frac{\partial }{\partial {\bf r}} \cdot D_{jnj} (x,x',x'',a;t,t')\cdot \frac{\partial }{\partial {\bf r}''}\nonumber     \qquad\qquad \qquad\\
-\frac{\partial }{\partial {\bf p}} \cdot D_{Fjn} (x,x',x'',a;t,t')\cdot \frac{\partial }{\partial {\bf r}'}
-\frac{\partial }{\partial {\bf p}} \cdot D_{Fnj} (x,x',x'',a;t,t')\cdot \frac{\partial }{\partial {\bf r}''}\nonumber \\
-\frac{\partial }{\partial {\bf r}} \cdot D_{jFn} (x,x',x'',a;t,t')\cdot \frac{\partial }{\partial {\bf p}'}
-\frac{\partial }{\partial {\bf r}} \cdot D_{jnF} (x,x',x'',a;t,t')\cdot \frac{\partial }{\partial {\bf p}''}\nonumber     \\
 +\frac{\partial }{\partial {\bf p}} \cdot D_{FFn} (x,x',x'',a;t,t')\cdot \frac{\partial }{\partial {\bf p}'}
+\frac{\partial }{\partial {\bf p}} \cdot D_{FnF} (x,x',x'',a;t,t')\cdot \frac{\partial }{\partial {\bf p}''}\bigg],\nonumber
\end{eqnarray}
It is remarkable that expression
$$\int dx'\int da\int_{-\infty}^{t}dt'e^{\epsilon(t'-t)}\phi_{nn}(x,x',a;t,t')f(a;t')\gamma (x';t')$$
in equation (\ref{Eq.23}) with (\ref{Eq.a5}) is the generalized collision integral of
Fokker-Planck type in the coordinate-momentum space.


\end{document}